\documentstyle[aps,prl,multicol,epsf]{revtex} 
\begin{document} 
\widetext
\title{\bf  On the pseudogap phase in high-$T_c$ superconductors}

\author{P.~ Devillard$^{(a)}$ and J. ~Ranninger$^{(b)}$}
\address{$^{(a)}$ Centre de physique th\'eorique de Marseille, C.P.T. 
Case 907, \\ Centre National de la
Recherche Scientifique, Luminy, 13288 Marseille C\'edex 9, France}

\address{$^{(b)}$ Centre de
Recherches sur les Tr\`es Basses Temp\'eratures, 
Laboratoire
Associ\'e \`a l'Universit\'e Joseph Fourier, 
\\ Centre National de la
Recherche Scientifique, BP 166, 38042, Grenoble 
C\'edex 9, France}

\date{\today} 
\maketitle 
\draft 
\begin{abstract} 
We describe the approach of the superconducting state as 
a sequence of cross-over phenomena. As the temperature is decreased, 
uncorrelated  pairing of the electrons leads to the 
opening of a pseudogap at $T_F^*$. Upon further lowering the 
temperature those electron pairs acquire well behaved itinerant 
features at $T_B^*$, leading to partial Meissner screening 
and Drude type behavior of the optical conductivity.
Further decrease of the temperature leads to their condensation and superconductivity at $T_c$. The analysis is done on the basis 
of the Boson-Fermion model in the cross-over regime between 2D and 3D.

\end{abstract}


\begin{multicols}{2}

\narrowtext

It is now generally accepted that the onset of superconductivity 
in high temperature superconductors (HTS) is
controlled by phase fluctuations\cite{Chakrav-94}, whilst a finite  
amplitude of the order parameter is expected to persist up to some 
$T^*$, which can be well above $T_c$. It 
has also been argued\cite{Emery-95} 
that the shrinking of the pseudogap phase (the interval 
$[T_c,T^*]$) with increased doping should be related to an increase of   
the superfluid density $n_s$ rather than of the coherence length. 
In that case, doping of HTS should not induce a cross-over between a Bose-Einstein condensation (BEC) of 
preformed pairs and a BCS state and hence the opening of the pseudogap 
should  be unrelated to the onset of superconducting fluctuations. 
These expectations seem to be confirmed by specific heat data\cite{Loram-98}, transient Meissner screening\cite{Corson-99} 
and Andreev spectroscopy\cite{Deutscher-99}. The picture  
evolves according to which a pseudogap opens up, exclusively driven 
by  amplitude fluctuations (dynamical pair-formation, uncorrelated in 
space and not necessarily related to 
incipient superconductivity). At some lower temperature 
 short-range  and short-time correlations of the electron pairs 
set in, eventually driving the system into a superconducting state.
 
From a theoretical point of view these are questions which can 
be addressed without having to resort 
to a specific microscopic mechanism for electron pairing and can be 
studied on such models as the generalized BCS hamiltonian, the 
negative U Hubbard model or the Boson-Fermion model (BFM). These
 phenomenological models\cite{Randeria-97} which explicitly 
incorporate pair 
correlations, capture a number of normal state properties of the HTS. 

In a classical BCS type superconductor long range phase coherence 
occurs as soon as short range amplitude correlations (in form of 
electron pairing) set in.  A simple  mean field treatment can perfectly  
describe this situation. On 
the contrary, in HTS amplitude and phase correlations are 
separated and a minimal treatment must contain the possibility of 
allowing for propagating modes of electron-pairs on a time scale  
of the order of the inverse zero temperature gap\cite{Tchernishyov-97}. 
We shall here explore the intricate relations between the onset of 
pairing of the electrons, their becoming itinerant and ultimately 
their condensation. In order to 
separate these various features in approaching the superconducting 
state we study them in view of a dimensional cross-over such as to 
capture part of the doping induced changes when going from the under 
into the overdoped regime. The dimensional cross-over will be 
controlled by the degree of anisotropy of the electron dispersion. 
The anisotropy in the electric transport 
coefficients\cite{resistivityexp} lends itself to such a picture.

This study is based on the BFM, a phenomenological 
model which is particularly suited for that purpose.
 The notion of rather long lived short 
range pair correlations is introduced explicitly into this model by 
assuming the presence of localized tightly bound electron pairs 
 which hybridize with pairs of itinerant electrons. 
We assume a system described by effective sites each of which 
can alternatively be occupied by either a bound electron-pair or a pair of 
itinerant electrons, uncorrelated with each other. For the present 
study we shall assume a homogeneous distribution of both the  bound 
electron-pairs and the itinerant electrons.
It is however feasible that
the distribution between those constituents occurs in form of a 
phase-separated phase with dynamically fluctuating regions of 
predominantly bound electron-pairs and regions with predominantly electrons, 
leading to ``stripe phases". 
Previous studies of the BFM\cite{Ranninger-95}, based on 
conserving diagrammatic approximations let one envisage a well 
defined finite temperature 
interval for the pseudogap phase in 1D and 2D, while for 3D  
such a phase\cite{Ren-98} should  be restricted to a 
very narrow temperature regime above $T_c$. This suggests 
that the nature of the pseudogap phase should depend on dimensionality 
and that the pair-fluctuations which control it have, depending 
on temperature, different characteristics as far as superconducting 
fluctuations are concerned. We expect that at least
two energy scales should be involved here, identifiable as two 
temperatures: $T^*_F \equiv T^*$ and $T^*_B \leq T^*_F$,
\begin{itemize}
\item 
$T^*_F$ corresponding to the opening of the pseudogap in the DOS of 
the electrons due to the onset of strong local correlations 
leading to electron pairing, uncorrelated in space.
\item
$T^*_B$ corresponding to those local pair states becoming well 
defined itinerant excitations due to the onset of temporal and 
spatial correlations.
\end{itemize}
We shall determine the dependence of $T^*_F$ and $T^*_B$
on dimensionality by monitoring the anisotropy of the ratio of the 
electron mass orthogonal to the basal plane to that within it, 
$\alpha= m_{\perp} / m_{\parallel}$ in the bare electron dispersion
\begin{eqnarray}
\varepsilon_{{\bf k}}=
\frac{1}{m_{\parallel}}\left(\frac{|{\bf k_{\parallel}}|}{\pi}\right)^2+
\frac{1}{m_{\perp}}\left(\frac{|{\bf k_{\perp}}|}{\pi}\right)^2 - 
\mu \nonumber
\end{eqnarray}
The BFM is then defined by the following Hamiltonian
\begin{eqnarray}
H=\sum_{{\bf k},\sigma} \varepsilon_{{\bf k}} 
c^{\dagger}_{{\bf k}\sigma}c_{{\bf k}\sigma} 
+E_0 \sum_{\bf q} b^{\dagger}_{\bf q} b_{\bf q} 
+v \sum_{\bf k} [b^{\dagger}_{\bf q} c_{{\bf k+q}\downarrow}
c_{{\bf -k}\uparrow} + h.c.] \nonumber ,
\label{eq2}
\end{eqnarray}
with $E_0=\Delta_B-2\mu$ and $v$ denoting the pair-exchange 
coupling constant.
$c_{{\bf k}\sigma}^{(\dagger)}$ denote fermionic operators for electrons 
with spin $\sigma$ and wavevector ${\bf k}$ and $b_{\bf q}^{(\dagger)}$ 
describe 
tightly bound electron pairs which will be considered as simple Bosons. 
The chemical potential $\mu$ is common 
to  Fermions and Bosons (up to a factor 2 for the Bosons) in 
order to guarantee charge conservation. We examine this model 
within the lowest order self-consistent conserving diagrammatic 
approximation for which the self-energies
for the Fermions and Bosons are given by:
\begin{eqnarray}
{\it Im}\Sigma_B({\bf q}, \omega) = 
-{v^2 \over 2\pi}
\int {d^3k \over (2\pi)^3}\int d\varepsilon {\it Im} 
G_F({\bf k}-{\bf q}, \omega-\varepsilon) \nonumber \\
{\it Im} G_F({\bf k},\varepsilon) 
\left( tanh({\varepsilon \over 2k_BT}) -
tanh({\varepsilon-\omega \over 2k_BT}) \right) \nonumber \\ 
{\it Im}\Sigma_F({\bf k}, \omega) =
{v^2 \over 2\pi}
\int {d^3q \over (2\pi)^3}\int d\varepsilon {\it Im} 
G_F({\bf k}-{\bf q},\varepsilon-\omega) \quad \nonumber \\
{\it Im} G_B({\bf q},\varepsilon) 
\left( tanh({\varepsilon-\omega \over 2k_BT}) - 
cotanh({\varepsilon \over 2k_BT})\right). \nonumber
\end{eqnarray}
$G_F({\bf k},\omega) = [\omega-\varepsilon_{{\bf k}}- 
{\it Re}\Sigma_F({\bf k}, \omega)-
i {\it Im}\Sigma_F({\bf k}, \omega)]^{-1}$ and  
$G_B({\bf q},\varepsilon)=[\omega-E_0 - 
{\it Re}\Sigma_B({\bf q}, \omega)-
i {\it Im}\Sigma_B({\bf q}, \omega)]^{-1}$ denote the fully 
renormalized Fermion and Boson Green's functions which 
have to be determined selfconsistently.
The above set of equations for the Green's functions is solved by 
a standard iterative procedure.  
The integration over the momenta ${\bf k_{\parallel}}$
is carried out by summing over a grid of 1025 points in the interval 
$[0,\pi]$. The integration over frequencies is carried out slightly 
above the real axis i.e., taking 
$\omega = {\it Re} \, \omega + i \, \eta$ 
with $\eta=0.01$ and by summation over 2048 real frequencies in 
the interval $[-2,2]$.Throughout the present work all energies 
are given in units of the bare basal plane
bandwidth $\widetilde {D}=8t_{\parallel}$.

The pseudogap features which result from the solution of the above 
set of equations depend strongly on dimensionality. In 1D  and 2D
the Fermionic self-energy is essentially determined by the term 
proportional to $cotanh$  which leads to the most divergent contribution.
For 3D the major contribution to the self-energy comes from the 
term proportional to $tanh$  and does not 
give rise to a noticeable pseudogap effect. A pseudogap is only 
seen as we lower the temperature and approach $T_c$ where the 
term proportional to $cotanh$ again becomes important. 

The various parameters entering our Hamiltonian are determined such 
as to reproduce certain robust features of HTS. Pinning down the number 
of Bosons $n_B=\langle b_i^+ b_i \rangle$ in this  
model is not free of ambiguities, given our poor understanding of 
the doping process in these materials. 
For a system such as, for instance, YBCO the number   
of doping induced bound elelctron-pairs could possibly be given by half the 
number of dopant ions $O_2^{2-}(1)$ in the chains, thus varying between 
$0$ and $0.5$ per effective site. The number of fermions  
$n_F=\sum_{\sigma} \langle c^{+}_{i\sigma}c_{i\sigma} \rangle$ is taken 
to be
equal to 1 if the boson-fermion exchange coupling were absent. We thus 
obtain a total number of charge carriers $n_{tot}=n_F+2n_B$ which 
should be contained between 1 and 2. As a representative example we 
choose  $n_{tot}=1.25$. In order to have $n_F \leq 1$ we 
fix the bosonic level as $\Delta_B=1.1$. 
In order to obtain values for $T^*$ of the 
order of a few hundred degrees K we choose $v=0.1$.

The pseudogap manifests itself as a dip which emerges in the DOS of the 
electrons close to the chemical potential  below a certain 
temperature $T_F^*$. Tracing the value of the DOS at this energy 
as a function of temperature permits to identify this characteristic 
temperature. The appearence of this pseudogap is linked, as we 
have discussed previously\cite{Ranninger-95}, to a breakdown of 
well defined single-electron excitations close to the 
Fermi surface. Concomitantly with this trend, two-electron states 
(local electron-pair resonances) emerge and 
upon lowering the temperature, acquire itinerant behaviour 
below a second  characteristic temperature $T_B^*$.  $T_B^*$ is 
defined by the condition (see Fig.1) that the imaginary part divided 
by the real part of the Boson self-energy is small, i.e., 
$\gamma_q^B(T)/\omega_q^B(T) \equiv \Gamma_B(q_{\parallel},
\omega_B(q_{\parallel},T),T)/
\omega_B(q_{\parallel}, T) \leq 0.1$ for small $q$ vectors such 
as $q_{\parallel}=0.1$. Below  $T_B^*$ we find 
$\omega_B(q_{\parallel}) = \hbar^2 q_{\parallel}^2/2m_B(T)$ 
for $q \leq 0.2$) where $m_B(T)$ denotes the temperature dependent 
mass of the itinerant two-particle excitations.

The two-electron states, becoming well defined as the temperature 
decreases, goes hand in 
hand with a significant increase of the distribution function 
for those states with small wave vectors. It is this feature 
which leads to a finite value of the amplitude of the order 
parameter in the normal state whilst at the same time phase 
coherence is totally absent. 
The temperature $T_c$ at which superconductivity sets in is 
determined by the condition that the two-electron 
excitations condense in a macroscopic quantum state. 
According to the Hugenholtz-Pines theorem, it is determined by 
$\Delta_B-2\mu-Re\Sigma_B(0,0)=0$. We plot in Fig.1 these three  
characteristic temperatures as a function of $\alpha$. Notice 
the opposite trends of $T_F^*$ and $T_c$, approaching each other 
as the isotropic limit is approached, ($\alpha=10)$. This is 
reminiscent of the experimental situation in 
cuprate HTS materials, as we go from the underdoped towards the 
optimally doped regime and might hence partly be related to such a 
dimensional cross-over.
\begin{figure}
\vspace{1mm}
\centerline{\epsfxsize=6cm \epsfbox{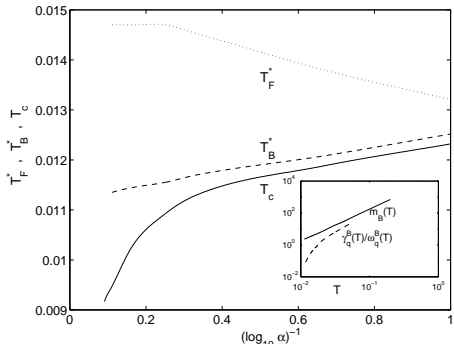}}
\caption{Variation of $T_F^*$, $T_B^*$ and $T_{c}$ as a 
function of decreasing  anisotropy. In the insert, the 
effective mass $m_B$ (in units of $m_\parallel(T)$) of the 
itinerant in-plane bosonic excitations as a function of 
temperature (solid line) and the ratio of their width divided 
by their real part; for $\alpha=10$.} 
\end{figure}
In order to track such cross-over behavior in the pseudogap phase, 
leading from uncorrelated fluctuating electron pairs to their 
itinerant behavior as the temperature is decreased, we now 
show how such features are related to  precursor Drude behavior 
of the optical conductivity and partial Meissner screening 
above $T_c$. We encounter here physics similar to that of the 2D 
non-interacting charged Bose  gas\cite{May-59},
where a macroscopic number of itinerant bosonic states with 
momenta ${\bf k} \leq 1/\xi$ ($\xi$ denoting the coherence length)
act together collectively in a way similar to the macroscopically 
occupied condensed states with ${\bf k}=0$ in the superconducting phase.
The optical conductivity $\sigma({\bf q}, \omega)$ and the 
diamagnetic susceptiblity $\chi({\bf q}, \omega)$ are given by 
the longitudinal, respectively 
transverse  part of the linear response to an external vector 
potential  ${\bf A}({\bf x},t)$  depending on space and time.
The resulting current is given by
\begin{eqnarray}
\langle J_i({\bf x},t)\rangle = {e^2 \over m_{\parallel}^2 c} 
\int d^3x dt K_{ij}({\bf x-x'}, t-t') A_j({\bf x'},t'),\nonumber \\
K_{ij}({\bf x}, t) = i\Theta(t)\langle[ j_i({\bf x}, t),j_j(0,0)]\rangle
-2m_{\parallel}\delta({\bf x})\delta(t)\delta_{ij}n^F(x) \nonumber
\end{eqnarray}
where $n^F(x)$ denotes the density of the Fermions and 
$j^F_i({\bf q}, t)= \sum_{\bf k} 2k_i 
c^{\dag}_{{\bf k-q/2}}c_{{\bf k+q/2}}$ the Fourier transform 
of their current density. Putting
\begin{eqnarray}
K_{ij}({\bf q}, \omega)= \delta_{ij} i \omega 
\sigma({\bf q}, \omega) -(q_iq_j-\delta_{ij}q^2)
\chi({\bf q}, \omega) \nonumber
\end{eqnarray}
permits to extract the optical conductivity and diamagnetic 
susceptibility. The kernel $K_{ij}({\bf x},t)$ is decomposed into two 
contributions. A first one is given by the simple bubble for 
the current autocorrelation function and neglecting vertex 
corrections. This is justified because of the strong incoherent 
contributions of $G_F$\cite{Ranninger-95}.
The second contribution to this kernel is evaluated in terms of 
the typical Aslamazov-Larkin diagram which takes into account the 
contributions of the itinerant Bosons. 
Due to the intrinsically localized 
nature of the Bosons there are no direct contributions of them to 
the electrical current. However since below $T_B^*$  those bosonic 
two-electron excitations become itinerant they will, according to 
this Aslamazov-Larkin mechanism, contribute to substantially enhance the 
conductivity. This is manifest in the 
dc resistivity (see insert of Fig.2) which clearly shows that upon 
decreasing the anisotropy ratio $\alpha$, the resistivity changes 
qualitatively from an upturn to a downturn as the temperature is 
lowered. Similarly, the effect of the pseudogap and 
the precursor to a macroscopic quantum state of 
the two-particle excitations can be tracked in the optical conductivity
(Fig.2). By inspection of Fig. 2 we notice that for frequencies above
$\simeq 0.01$ the optical conductivity drops as the temperature is 
decreased below $T =0.03$  (indicative of a remnance of the 
pseudogap in the DOS of the single-electron states), while for 
frequencies below $\simeq 0.01$ the optical conductivity for those 
temperatures increases (indicative of an emerging precursor ``Drude" 
component as we approach $T_c$). The crossover temperature at 
$T=\simeq 0.01$ is identified as the temperature $T^*_B$ where 
the electron-pairs become itinerant. As the temperature is lowered, spectral 
weight is shifted downwards from the frequency regime 
$[0.01 \leq \omega \leq 0.04]$ thus enhancing this ``Drude" component 
below $0.01$. Such a redistribution of spectral 
weight is in fact observed\cite{Basov-97}. For frequencies below 0.001 
($\simeq 240 \, GHz$, taking $\widetilde{D}=1 \, eV$) the optical 
conductivity increases substantially as one approaches $T_c$.
 
Similarly, the real part of the diamagnetic susceptibility 
$\chi'({\bf q}, \omega)$ for small wavevectors 
($q \leq 1/\xi$, $\xi=1/\sqrt{2m_B(T)\omega_0^B}$ denoting the 
coherence length) 
and frequencies is negative and small for high temperatures. 
It is given by the usual Landau diamagnetism of the electrons arising 
from the first contribution to the kernel $K_{ij}$.
For temperatures below $T^*_B$ (which is $0.01163$ for a representative
 example of $\alpha = 10^3$) we calculate $\chi^{\prime}(q,\omega)$,
not taking into account the phase fluctuations. Close to $T_c$ $(= 0.01127$
for $\alpha = 10^3)$  $\chi^{\prime}(q,0)$ however is expected  
to diverge as $1/\xi_{KTB}^2$, where $\xi_{KTB}^2$ is the 
Kosterlitz-Thouless-Berezinskii 
coherence length for phase fluctuations. Our calculated 
$\chi^{\prime}(q,\omega)$ increases significantly (roughly 
by a factor 50) over the Landau diamagnetism (see Fig. 3). 
It is given by the Aslamazov-Larkin contribution of $K_{ij}$
i.e., $\chi^{\prime}(0,0) \simeq 
- a v^4 ({e^2 \over 24 \pi m_B (T)c^2} {k_B T \over \omega_0 ^B})$,
$a$ being a numerical factor of order unity. The expression in
the bracket corresponds to that known from the 2D 
Bose gas\cite{May-59}. It is this contribution which leads to almost
complete Meissner screening.
 A characteristic frequency 
scale thus emerges below which phase  uncorrelated amplitude 
fluctuations of the electron-pairs behave 
in some respects as condensed electron-pairs in the superconducting 
state. For frequencies $\omega \geq 10^{-5}$ we find the usual Landau 
diamagnetism but for  $\omega \leq 10^{-5}$ we notice 
a saturation of $\chi'({\bf q}, \omega)$ which decreases as 
the temperature increases. It is in this low frequency regime 
that we can extract from the diamagnetic susceptibility an almost 
complete Meissner screening for strong anisotropies. Its physical 
meaningfulness arises 
from the fact that for such a system there is phase locking over a 
given finite distance which is controlled essentially by the thermal 
wavelength of the itinerant Bosons\cite{Zhu-94}. 
\begin{figure}
\vspace{1mm}
\centerline{\epsfxsize=8cm \epsfbox{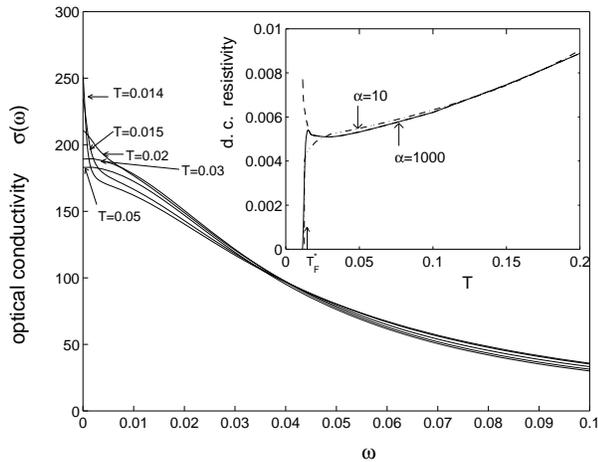}}
\caption{The in-plane optical conductivity (arbitrary units) 
for a set of different temperatures and $\alpha=10$. The 
inset illustrates the d.c. resistivity as a function of $T$, 
for $\alpha=10$ including Aslamazov-Larkin contributions 
corrections (dashed dotted line) and  for $\alpha=1000$ including 
(solid line), respectively excluding them (dashed line). 
The arrow on the $T$ axes indicates $T_F^*$ for $\alpha=1000$.}
\end{figure}
In this Letter we have examined the effect of pure amplitude 
fluctuations in the pseudogap phase of HTS as the dimensionality 
of the system varies between quasi-2D and quasi-3D. We conclude 
that the doping induced cross-over between the under- and the 
over-doped regime might at least partly be related to such 
a dimensional cross-over, clearly indicating a shrinking of the 
temperature regime for the pseudogap phase as the optimally doped 
limit is approached. We identified two characteristic 
temperatures in the pseudogap phase. The first one, $T_F^*$, 
corresponding to the opening of the pseudogap due to the onset of local 
uncorrelated electron-pair correlations. The second one, $T_B^*$, 
corresponding to the onset of itinerant behavior of those electron-pairs, 
below which we expect partial Meissner screening  (very similar to that  
expected for the 2D non-interacting charged Bose gas, i.e., in the 
absence of any phase  fluctuations) and an optical conductivity 
showing Drude behavior. 

\begin{figure}
\vspace{1mm}
\centerline{\epsfxsize=8cm \epsfbox{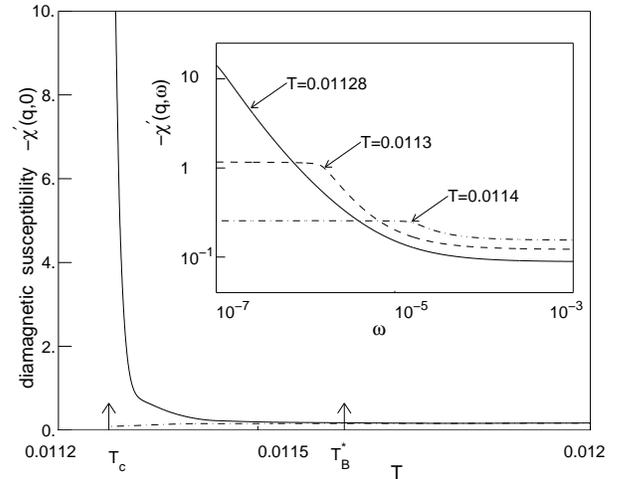}}
\caption{Real part of the in-plane diamagnetic susceptibility 
for small  $q$ and frequency $\omega=0$ as a 
function of temperature. In the insert, its frequency 
dependence for different temperatures. $\alpha=1000$.}
\end{figure}

We would like to acknowledge helpful discussions with B. K. 
Chakraverty, T. Domanski, K. Matho and A. Romano. This research 
was supported in part by a European network contract ERBCHRXCT940438.

\end{multicols}

\end{document}